\documentclass[a4paper,11pt]{article}
\usepackage{pos}
\usepackage{subcaption}
\usepackage{slashed}            % Feynman slash notation
\usepackage{listings}		% Code
\usepackage{braket}
\usepackage{soul}                  % For highlighting (\hl{text to be highlighted})

\usepackage{multirow}   % table
\usepackage{hhline}     % table: double line

\usepackage[compact]{titlesec} % To adjust spacings
\titlespacing{\section}{0pt}{10pt}{3pt} % this reduces space between (sub)sections to 0pt, for example

\newcommand{\code}[1]{\lstinline|#1|}  % For inline code

\definecolor{mygreen}{rgb}{0,0.6,0} 
\definecolor{mygray}{rgb}{0.5,0.5,0.5} 
\definecolor{mymauve}{rgb}{0.58,0,0.82} 

\title{Improving HISQ propagator solves using deflation}
\author*[a]{Leon Hostetler}
\author[b]{M. A. Clark}
\author[c]{Carleton DeTar}
\author[a]{Steven Gottlieb}
\author[b]{Evan Weinberg}

\affiliation[a]{Department of Physics, Indiana University,\\
  Bloomington, Indiana 47405, USA}
  
\affiliation[b]{NVIDIA Corporation,\\
  Santa Clara, California 95051, USA}

\affiliation[c]{Department of Physics \& Astronomy, The University of Utah,\\
Salt Lake City, Utah 84112, USA}

\emailAdd{leonhost@iu.edu}
\emailAdd{mclark@nvidia.com}
\emailAdd{detar@physics.utah.edu}
\emailAdd{sg@iu.edu}
\emailAdd{eweinberg@nvidia.com}

\abstract{
Typically, the conjugate gradient (CG) algorithm employs mixed precision and even-odd preconditioning to compute propagators for highly improved staggered quarks (HISQ). This approach suffers from critical slowing down as the light quark mass is decreased to its physical value. Multigrid is one alternative to combat critical slowing down; however, it involves setup costs that are not always easy to amortize. We consider deflation, which can also remove critical slowing down, but incurs its own setup cost to compute eigenvectors. Results using the MILC and QUDA software libraries to generate eigenvectors and to perform deflated solves on lattices up to $144^3 \times 288$ (with lattice spacing 0.04 fm) and with a range of quark masses from the physical strange down to the physical light quark values will be presented. We compare with CG and comment on deflation versus multigrid.

}

\FullConference{The 41st International Symposium on Lattice Field Theory (LATTICE2024)\\
 28 July - 3 August 2024\\
Liverpool, UK\\}

\begin{document}
\maketitle

%%%%%%%%%%%%%%%%%%%%%%%%%%%%%%%%%%
\section{Introduction}

In lattice QCD calculations, solving the Dirac equation is the dominant computational bottleneck since it has to be done repeatedly both in the hybrid Monte Carlo algorithm used to generate configurations and in the propagator solves used to compute correlators. The solves are typically done using an iterative Krylov subspace method such as the conjugate gradient (CG) algorithm, and the computational effort required is dependent on the condition number of the system. To get more precise results from lattice QCD, we are driven to use finer lattices. However, as our lattice spacing becomes finer, the bare light quark mass corresponding to physical hadrons becomes smaller, and since the smallest eigenvalue of the Dirac matrix is essentially the quark mass, we end up confronting a diverging condition number. This is the problem of ``critical slowing down,'' and it hits us at both ends---configuration generation and propagator solves---of the typical lattice QCD calculation.

One approach is to use eigenvector deflation \cite{Forcrand_1996, Stathopoulos_2010, Davies_2018, Romero_2020}, which solves for the low modes exactly, leaving the CG solver to work on the easier high mode part of the solution. Deflation does not eliminate critical slowing down, but only shifts it from the CG solve to the eigensolve. Nonetheless, in common lattice QCD workflows, this can be very beneficial. Multi-grid (MG) algorithms are another approach, and they work well to eliminate critical slowing down for Wilson \cite{Brannick_2008, babich_2010, osborn_2010}, twisted mass \cite{Frommer_2014, Richtmann_2022}, and domain wall \cite{Cohen_2012, boyle_2014, Brower_2020} fermions. An efficient MG method for staggered fermions has been more elusive, however, significant recent progress has been made using a K{\"a}hler-Dirac preconditioned MG algorithm \cite{Brower_2018, Ayyar_2023}.

The cost of computing an eigenvector is proportional to the lattice volume $V$, and because the density of low modes increases with volume, the number of eigenvectors needed for deflation scales with the volume. Because of the $V^2$ cost of deflation, it is not a long-term solution for lattice QCD. In the long run, MG methods will be needed to truly eliminate critical slowing down. For propagator solves with highly-improved staggered quarks (HISQ) using current state of the art algorithms, it is not clear whether MG or deflated CG performs better on current lattices. This is largely due to limited numerical results in the literature for HISQ deflation performance \cite{Davies_2018} and none for the largest lattice sizes in use. This is the gap that the present work attempts to fill. We use the MILC and QUDA \cite{Clark_2010,Babich_2011} software libraries to investigate the performance of deflated CG for a range of quark masses and lattice sizes.

%%%%%%%%%%%%%%%%%%%%%%%%%%%%%%%%%%
\section{Critical Slowing Down in HISQ Propagator Solves}

The conventional approach to compute HISQ propagators is to use a mixed precision CG algorithm. Typically, this is even-odd preconditioned CG---without deflation---performed on the normal equations. Quark propagators $\psi$ are computed by solving the linear equation $M\psi=\eta$, where $\eta$ is a source field, $M\equiv \slashed{D}+2m$, $D$ is the anti-Hermitian HISQ Dirac matrix, and the factor of 2 in front of the bare quark mass $m$ is a historical MILC convention. The CG algorithm, in its original form, works only for positive definite systems. The HISQ operator $M$ is not positive definite, but the squared operator $M^\dagger M$ is, so in practice, we solve the normal equation
\begin{equation}
\label{normaleq}
M^\dagger M\psi = M^\dagger\eta.
\end{equation}
Then $\psi=(M^\dagger M)^{-1}M^\dagger\eta = M^{-1}\eta$. In the following, we will refer to the conventional even-odd preconditioned conjugate gradient approach simply as ``CG'', and it will be the baseline we use for performance comparison with deflated CG.

The number of CG iterations needed to reach the solution (throughout this article, the stopping criterion is a true residual $<10^{-8}$) depends on the condition number
\begin{equation}
\kappa = \frac{\lambda_{max}}{\lambda_{\min}},
\end{equation}
of the system being solved. Here $\lambda$ are the eigenvalues of the squared massive operator $M^\dagger M$. Empirically, we find $\lambda_{max} \approx 23$ for HISQ. The smallest eigenvalue of the squared massive operator is
\begin{equation}
\lambda_{min} = \epsilon_{min} + (2am_q)^2,
\end{equation}
where $\epsilon_{min}$ is the smallest eigenvalue of the squared massless operator $\slashed{D}^\dagger\slashed{D}$. Since $\epsilon_{min}\approx 0$, the condition number and the number of CG iterations diverge as the quark mass $am_q$ is decreased.

The ensemble parameters used in this study are detailed in Table~\ref{condnum}. The first columns give the lattice size, the approximate lattice spacing $a$, the physical spatial extent of the lattice, the bare light quark mass corresponding to physical pions, and the strange quark mass. The column $\epsilon_{min}$ gives the smallest eigenvalue of the massless operator $\slashed{D}^\dagger \slashed{D}$ as reported by the \code{staggered_eigensolve_test} application from QUDA, and $\kappa$ gives the approximate condition number of the operator $M^\dagger M$. Notice that the ensembles approach the continuum limit in lattice spacing at fixed physical volume of 5.76 fm.

\setlength{\tabcolsep}{0.5em} % for the horizontal padding
\renewcommand{\arraystretch}{1.1}
\begin{table*}
\centering
\begin{tabular}{|c|c|c|c|c|c|c|} \hline
lattice & $\sim a$ & physical size & $am_\ell$ & $am_s$ & $\epsilon_{min}$ & $\kappa$ \\ \hline
$48^3 \times 64$ & 0.12 & 5.76 fm & 0.001907 & 0.05252 & $2.4 \times 10^{-10}$ & $1.6 \times 10^6$\\ \hline
$64^3 \times 96$ & 0.09 & 5.76 fm & 0.001200 & 0.03630 & $4.1 \times 10^{-10}$ & $4.0 \times 10^6$\\ \hline
$96^3 \times 192$ & 0.06 & 5.76 fm & 0.000800 & 0.02200 & $9.9 \times 10^{-11}$ & $9.0 \times 10^6$ \\ \hline
$144^3 \times 288$ & 0.04 & 5.76 fm & 0.000569 & 0.01555 & $1.4 \times 10^{-11}$ & $1.8 \times 10^7$ \\ \hline
\end{tabular}
\caption{Details of the ensembles used in this study---the lattice size, approximate lattice spacing $a$, physical spatial extent of the lattice, the physical light quark mass, the strange quark mass, the smallest eigenvalue of the massless operator $\slashed{D}^\dagger \slashed{D}$ for the particular configuration studied, and the approximate condition number of the massive operator $M^\dagger M$ at the physical light quark mass.}
\label{condnum}
\end{table*}

\begin{figure}
\centering
\begin{subfigure}{.49\textwidth}
  \centering
  \includegraphics[width=\linewidth]{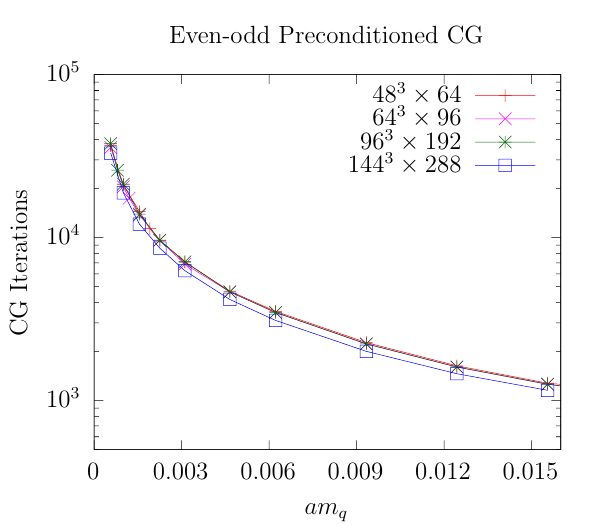}
\end{subfigure}
\begin{subfigure}{.49\textwidth}
  \centering
  \includegraphics[width=\linewidth]{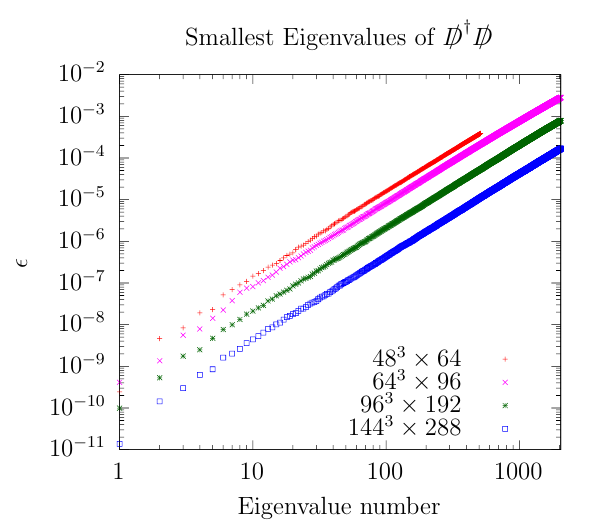}
\end{subfigure}
\caption{\textbf{(Left)} The number of CG iterations required to reach the solution versus the bare quark mass for several lattice sizes. \textbf{(Right)} The smallest eigenvalues of the squared massless Dirac operator $\slashed{D}^\dagger\slashed{D}$ for several lattice sizes. The ensembles approach the continuum limit in lattice spacing at fixed physical volume.}
\label{fig_critical_slowing_down}
\end{figure}

Critical slowing down can be visualized by plotting the number of CG iterations required to reach the solution as a function of the bare quark mass. See the left panel in Fig.~\ref{fig_critical_slowing_down}, and note the log scale on the vertical axis. Here it takes approximately thirty times more work to compute a light quark propagator than it does to compute a heavy quark propagator. Notice that the curves are roughly independent of the lattice size. As shown in the right panel of Fig.~\ref{fig_critical_slowing_down}, the smallest eigenvalues $\epsilon_{min}$ of $\slashed{D}^\dagger\slashed{D}$ range from about $10^{-11}$ to $10^{-9}$ for the different lattice sizes considered. So $\epsilon_{min} \approx 0$, which implies the condition number $\kappa \approx 23/(2am_q)^2$ approximately depends only on the quark mass (and not on the lattice size). What \textit{does} change with lattice size is the value of the bare quark mass needed to produce hadrons at the physical point. As we go to finer lattices, the physical light quark mass $am_{\ell,phys}$ gets smaller, and the condition number worsens.

For these experiments, eigenvectors of the massless squared Dirac operator $\slashed{D}^\dagger \slashed{D}$ are generated using the thick restarted Lanczos algorithm via QUDA's \code{staggered_eigensolve_test} application. Propagators $\psi$ are computed by solving Eq.~(\ref{normaleq}) using MILC's \code{ks_spectrum} application with the deflation and CG offloaded to QUDA's solver. In MILC, even-odd preconditioning (Schur decomposition) is performed $y=M^\dagger \eta$, and then the even site source $y_e$ is passed to QUDA, which performs the CG solve and returns the even-site solution $\psi_e=(M^\dagger M)^{-1} y_e$. The odd site solution $\psi_o = m(D_{oe}\psi_e + \eta_o)/2$ is reconstructed in MILC and passed to the QUDA solver which polishes it using one or more CG iterations $\psi_o = (M^\dagger M)^{-1} y_o$. This procedure is repeated for each of the three color charges.

The eigenvectors ${v_i}$ corresponding to the smallest eigenvalues are projected onto the source vector
\begin{equation}
x = \sum_i {v_i} \frac{1}{\lambda_i} v_i^\dagger \eta.
\end{equation}
Then a deflated CG algorithm is effectively achieved by passing $x$ as the initial guess to the CG solver. Then the solver only has to deal with the high modes which converge more quickly. The critical slowing down is effectively shifted from the CG solve to the eigensolve, and so the penalty is paid only once per gauge configuration instead of once per solve. Thus, deflation may be advantageous in the (fairly common) situation in which many propagators are computed for each gauge configuration---allowing the eigensolve cost to be amortized over a large number of propagator solves. We target that situation here by focusing only on the solve time and assuming that the eigensolve cost is made relatively negligible by the number of solves done per configuration.

%%%%%%%%%%%%%%%%%%%%%%%%%%%%%%%%%%
\section{Multi-deflation with Sloppy Eigenvectors}

\begin{figure}
\centering
\begin{subfigure}{.5\textwidth}
  \centering
  \includegraphics[width=\linewidth]{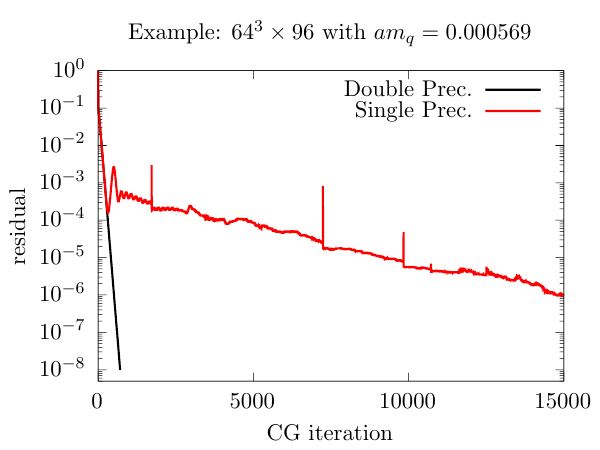}
  %\caption{A subfigure}
  %\label{fig:sub1}
\end{subfigure}%
\begin{subfigure}{.5\textwidth}
  \centering
  \includegraphics[width=\linewidth]{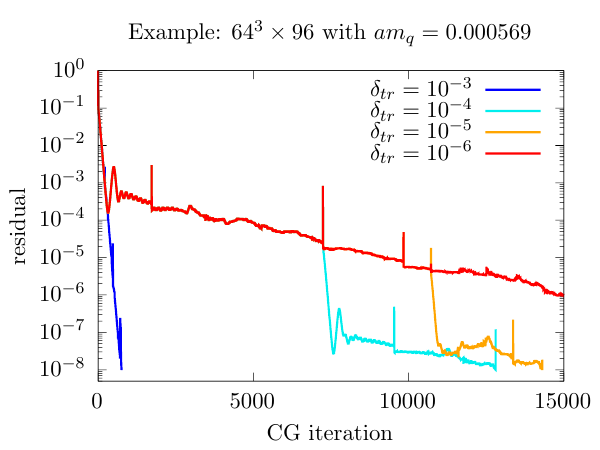}
  %\caption{A subfigure}
  %\label{fig:sub2}
\end{subfigure}
\caption{\textbf{(Left)} The iterated residual versus CG iteration number for (a single) deflation on a $64^3\times 96$ lattice with 2048 eigenvectors and at a lighter-than-physical quark mass of $0.000569$. The black curve shows the result when using double precision eigenvectors, and the red curve shows the result when using single precision eigenvectors. In the double precision case, the residual reaches $10^{-8}$ in 708 CG iterations, whereas in the single precision case, the solve stalls, and it takes $>>15,000$ iterations to reach the same residual. \textbf{(Right)} Now with multi-deflation using only single precision eigenvectors. Redeflation is triggered whenever the residual drops by a factor of $1/\delta_{tr}$. In this example, the number of CG iterations needed to reduce the residual to $10^{-8}$ is 755, 12804, 14305, and 18621 for $\delta_{tr} = 10^{-3}$, $10^{-4}$, $10^{-5}$, and $10^{-6}$ respectively. The key point is that by using single precision eigenvectors, one can achieve performance similar to double precision by redeflating whenever the residual drops by some empirically chosen factor.}
\label{fig_stagnation}
\end{figure}

The size of eigenvectors is a significant disadvantage for the deflation method in terms of the disk space needed to store them, the memory needed during generation and deflation, and even the IO time\footnote{The IO time can be significantly reduced by using \code{PARTFILE} format for the eigenvectors such that each MPI rank reads and writes to its own file.} spent reading the eigenvectors. Naively, the size of double precision eigenvectors is $V \times N \times 48$ bytes, where $V$ is the size of the lattice, and $N$ is the number of eigenvectors. For 2K eigenvectors of a $144^3\times 288$ lattice, this comes to more than 80~TB. However, with single parity format\footnote{See, e.g., Eq.~(6) and related text in Ref.~\cite{Davies_2018} .}, there is no need to ever compute or store the odd part of the eigenvectors, so the size can immediately be halved. This reduces the $144^3\times 288$ eigenvectors to 40~TB. Another factor of two can be gained by using single precision eigenvectors---reducing the size to 20~TB per gauge configuration. However, as we see in the left panel of Fig.~\ref{fig_stagnation}, using ``sloppy'' single precision eigenvectors has a detrimental stalling effect on deflation.

In QUDA's solver, a ``reliable update'' \cite{Clark_2010} is triggered when the residual drops by a factor of $1/\delta_{rd}$ since the last reliable update. During a reliable update the iterated residual is replaced by the true residual in a process that corrects the residual without doing a full restart of the CG. In our case, we used $\delta_{rd}=0.1$ (this is \code{reliable_delta=0.1} in QUDA), so a reliable update is performed whenever the residual drops by a factor of 10. To combat the problem of the residual stalling with sloppy eigenvectors, we experimented with periodically re-deflating the system during the CG solve. A redeflate was triggered whenever the residual dropped by a factor of $1/\delta_{tr}$ since the last deflation. As shown in the right panel of Fig.~\ref{fig_stagnation}, for the $64^3 \times 96$ lattice, with $\delta_{tr}=10^{-3}$ (this is \code{tol_restart=1e-3} in QUDA) we were able to eliminate the stalling effect and achieve an iteration count similar to that with double precision eigenvectors. In this figure, the blue residual curve (which behaves very similarly to the residual curve in the double precision case), is obtained with an initial deflation before the CG, followed by a secondary deflation when the residual drops to $10^{-3}$, and a tertiary deflation when the residual drops to $10^{-6}$. The choice of $\delta_{tr}$ was made empirically and varied a little with lattice size. A similar multi-deflation approach for handling approximate eigenvectors was used in Ref.~\cite{Stathopoulos_2009}. We found the cost of doing a few additional deflations to be relatively small. The primary benefit of the multi-deflation method is the memory footprint reduction achieved by working with single instead of double precision eigenvectors, which allows the problem to be solved in roughly the same amount of time on half as many nodes. Since fewer nodes means less communication, there is also a minor but real speedup of the solver kernels.

\setlength{\tabcolsep}{0.5em} % for the horizontal padding
\renewcommand{\arraystretch}{1.0}
\begin{table*}
\centering
\begin{tabular}{cccccccccc} \hhline{==========}
\multirow{1}{*}{} & \multicolumn{2}{c}{} & \multicolumn{3}{c}{$m_q=m_s$} & \multicolumn{1}{c}{} & \multicolumn{3}{c}{$m_q=m_{\ell}$} \\
\cline{4-6} \cline{8-10}
lattice           & EVs  &$\qquad$ & iters & time (s) & speedup &$\qquad$ & iters & time (s) & speedup \\ \hline
$48^3\times 64$   & 0    && 372 & 0.59 & --- && 11340 & 12.4 & --- \\
$48^3\times 64$   & 512  && 364 & 0.49 & 1.2 && 2030 & 2.70 & 4.6 \\
$48^3\times 64$   & 1024 && 347 & 0.60 & 0.98 && 988 & 1.18 & 10.5 \\
$64^3\times 96$   & 0    && 530 & 0.71 & --- && 17468 & 20.7 & --- \\ 
$64^3\times 96$   & 1024 && 490 & 1.21 & 0.59 && 1351 & 2.27 & 9.1 \\ 
$64^3\times 96$   & 2048 && 422 & 0.89 & 0.80 && 704 & 1.42 & 14.6 \\
$96^3\times 192$  & 0    && 879 & 1.36 & ---   && 25921 & 39.3 & ---  \\ 
$96^3\times 192$  & 1024 && 827 & 1.57 & 0.87 && 2774  & 5.48 & 7.2 \\ 
$96^3\times 192$  & 2048 && 730 & 1.77 & 0.77 && 1382  & 3.98 & 9.9 \\
$144^3\times 288$ & 0    && 1156 & 1.64 & --- && 33003 & 47.8 & --- \\ 
$144^3\times 288$ & 1024 && 1123 & 1.67 & 0.98 && 5593 & 8.72 & 5.5 \\ 
$144^3\times 288$ & 2048 && 1058 & 1.77 & 0.93 && 2909 & 4.79 & 10.0 \\ \hhline{==========}
\end{tabular}
\caption{Performance data from Perlmutter for solves at the strange quark mass $m_s$ and the light quark mass $m_{\ell}$ on four different lattice sizes and with different numbers of eigenvectors (EVs) for the deflation. For both quark masses, the number of CG iterations (iters) and the time to solution averaged over multiple solves are given. Speedups are calculated relative to the undeflated case with EVs $=0$.}
\label{speedups_table}
\end{table*}

\begin{figure}
\centering
\begin{subfigure}{.5\textwidth}
  \centering
  \includegraphics[width=\linewidth]{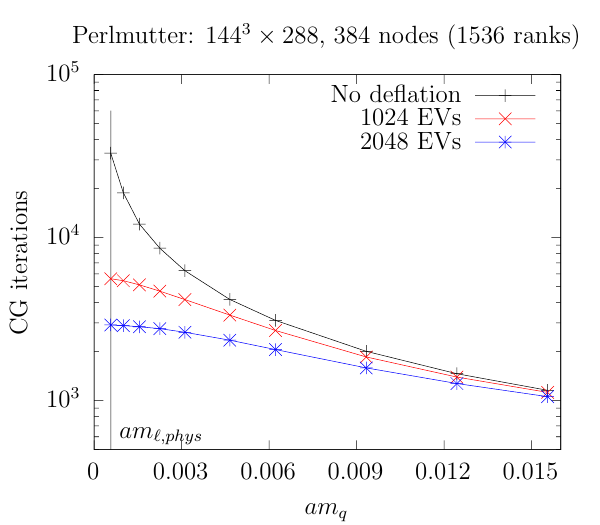}
  %\caption{A subfigure}
  %\label{fig:sub1}
\end{subfigure}%
\begin{subfigure}{.5\textwidth}
  \centering
  \includegraphics[width=\linewidth]{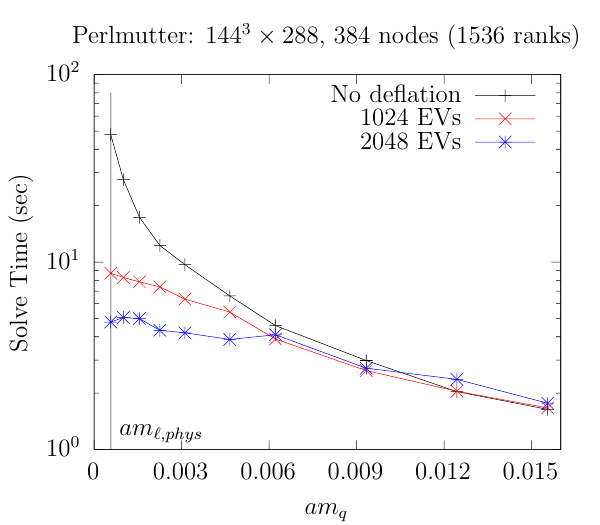}
  %\caption{A subfigure}
  %\label{fig:sub2}
\end{subfigure}
\caption{\textbf{(Left)} CG iterations versus bare quark mass for the $144^3\times 288$ system running on 1536 GPUs on Perlmutter. At the physical light quark mass $am_{\ell,phys}$, the undeflated case required on average 33,003 iterations per solve versus 2,909 for deflation with 2K eigenvectors. This implies a potential speedup of $\sim 11\times$ for the deflated case. \textbf{(Right)} Solve time versus bare quark mass for the same cases. At $am_{\ell,phys}$, the average solve time was 47.8s for the undeflated case versus 4.79s for the deflated case with 2K eigenvectors---an actual solve time speedup of 10x.}
\label{fig_results}
\end{figure}

To test the performance of multi-deflation with sloppy eigenvectors, we did runs with six solves (2 propagators $\times$ 3 colors) with a CG stopping criterion of $|r|/|b|<10^{-8}$, where $r$ is the residual and $b$ is the source vector. We noted the CG iterations and the solve time. The first solve is slower due to various setups and memory allocations, so we took the average of the last five solves when reporting the time to solution. This was repeated for various bare quark masses, lattice sizes, and numbers of eigenvectors used in the deflation. During the conference, we showed performance results from experiments which were performed on Frontier. However, we have since noticed a performance regression on Frontier where a large memory operation (e.g., computing the eigenvalues from 2K eigenvectors) caused subsequent \code{MPI_Allreduce} calls to run significantly slower. The result was a significant performance penalty for deflation. We did not see this issue on Perlmutter, hence, in these proceedings we show updated performance results taken on Perlmutter. Performance data for solves at the strange and light quark masses are given in Table~\ref{speedups_table} for different lattice sizes and using different numbers of eigenvectors (EVs) for the deflation. For both quark masses, the number of CG iterations (iters) and the time to solution averaged over multiple solves are given. Speedups are calculated relative to the undeflated case with EVs $=0$. The results show that deflated CG typically performs worse at the strange quark mass but shows significant speedups at the light quark mass. For the largest lattice, more detailed results are shown in Fig.~\ref{fig_results}. The left panel in the figure plots the CG iterations versus bare quark mass for conventional (undeflated) CG and for deflated CG using 1024 and 2048 eigenvectors. At the physical light quark mass $am_{\ell,phys}=0.000569$, the undeflated case required an average of 33,003 iterations per solve in order to reach a residual $<10^{-8}$ on each of the six solves. For the deflated case with 2048 eigenvectors, it only required 2,909 iterations. This ratio implies a potential speedup of about 11$\times$ for deflation with 2048 eigenvectors. In this example, the slope shows that critical slowing down is not completely eliminated even when deflating with 2048 eigenvectors. The performance could be improved by increasing the number of eigenvectors, but this would, of course, require running the problem on even more nodes. In the right panel of the figure, we show the average time to solution for the six propagator solves. This excludes setup times such as the eigensolve, loading the eigenvectors from disk, allocating memory, etc. At $am_{\ell,phys}$, the average solve time was 47.8s for the undeflated case versus 4.79s for the deflated case with 2048 eigenvectors---an actual solve time speedup of 10$\times$.

For the comparison shown in Fig.~\ref{fig_results}, and for a given lattice size listed in Table~\ref{speedups_table}, all three cases were run on the same resources. For the $144^3\times 288$ lattice, we used 384 Perlmutter nodes (1536 NVIDIA A100 GPUs each with 40 GB of memory). The memory-intensive deflated case with 2048 eigenvectors dictated the resources needed for this job. The less intensive 1024-eigenvector case could be run on about half as many nodes, and the undeflated case could run on many fewer nodes. Given that increasing the number of nodes introduces additional communication overhead, one could argue that this is an unfair representation of the relative cost in particular for the undeflated case. In addition to considering the speedup given identical resources, as in Table~\ref{speedups_table} and Fig.~\ref{fig_results}, it may be useful to consider also the cost efficiency as determined by minimizing the node-seconds per solve. With 384 Perlmutter nodes, the undeflated case used 18,355 node-seconds per solve at the light quark mass, and deflation with 2048 eigenvectors used 1839 node-seconds per solve. The undeflated case, requiring much less memory, was run on as few as 27 nodes where it used 7914 node-seconds per solve---much more cost efficient than running undeflated on 384 nodes, but still more than a factor of 4 less cost efficient than using deflation.

%%%%%%%%%%%%%%%%%%%%%%%%%%%%%%%%%%
\section{Summary and Outlook}

Critical slowing down, the phenomenon in which the condition number of the Dirac operator diverges as the quark mass goes to zero, is a significant bottleneck for lattice QCD calculations. Eigenvector deflation and multi-grid (MG) methods are two ways in which critical slowing down can be mitigated or eliminated. In these proceedings, we present results of our numerical experiments using deflation for propagator solves for highly improved staggered quarks (HISQ) using the MILC and QUDA software libraries. A primary motivation was to establish a baseline for comparison with HISQ MG experiments. We show a factor of 10 speedup of the time to solution on the $144^3\times 288$ lattice at the physical light quark mass over conventional even-odd preconditioned conjugate gradient (CG) by using deflation with 2048 eigenvectors. This speedup is comparable to recent results using MG preconditioned GCR \cite{Ayyar_2023}, where they also reported a 10$\times$ speedup for the same quark mass and lattice size. We also show that single (instead of the usual double) precision eigenvectors can be used, which halves the disk space and memory footprints, provided that the system is periodically re-deflated during the solve.

Further improvements are expected to come from performing the deflation on multiple right-hand sides concurrently during a batched solve, which would increase the arithmetic intensity and give greater floating point throughput. That functionality, which is now present in QUDA \cite{Clark_2024, Weinberg_2024}, is being actively pursued for deflation. Optimizing the eigensolve by, e.g., using block TRLM and the disk usage by, e.g., using eigenvector compression \cite{Clark_2018}, are also being pursued. Finally, we are continuing to improve the HISQ MG algorithm present in QUDA, including recent work on multiple right-hand-side support. The full set of above-mentioned work, including an investigation of both setup and solver times, will enable a robust comparison between deflation and MG methods for real lattice QCD workflows on modern exascale-ready ensembles.

%%%%%%%%%%%%%%%%%%%%%%%%%%%%%%%%%%
\acknowledgments

L.~H. thanks Peter Boyle and Michael Lynch for helpful discussions. L.~H. was supported by the SciDAC: H.E.P., LAB 22-2580 program under the U.S. Department of Energy, Office of Science and by the National Science Foundation (NSF) under Grant No. 2139536. Early exploration for this project was performed on computational resources provided by Michigan State University's ICER. Further development was done on resources provided by the OLCF under the ALCC program and by NERSC under the ERCAP program. C. D. was supported by NSF grant PHY23-10571.

\bibliographystyle{JHEP}
\bibliography{pos24}

\providecommand{\href}[2]{#2}\begingroup\raggedright\begin{thebibliography}{10}

\bibitem{Forcrand_1996}
P.~{de Forcrand}, \emph{Progress on lattice qcd algorithms},
  \href{https://doi.org/https://doi.org/10.1016/0920-5632(96)00047-3}{\emph{Nuclear
  Physics B - Proceedings Supplements} {\bfseries 47} (1996) 228}.

\bibitem{Stathopoulos_2010}
A.~Stathopoulos and K.~Orginos, \emph{Computing and deflating eigenvalues while
  solving multiple right-hand side linear systems with an application to
  quantum chromodynamics}, \href{https://doi.org/10.1137/080725532}{\emph{SIAM
  Journal on Scientific Computing} {\bfseries 32} (2010) 439}
  [\href{https://arxiv.org/abs/0707.0131}{{\ttfamily 0707.0131}}].

\bibitem{Davies_2018}
C.~Davies, C.~DeTar, C.~McNeile and A.~Vaquero, \emph{Numerical experiments
  using deflation with the hisq action},
  \href{https://doi.org/10.1051/epjconf/201817514016}{\emph{EPJ Web of
  Conferences} {\bfseries 175} (2018) 14016}
  [\href{https://arxiv.org/abs/1710.07219}{{\ttfamily 1710.07219}}].

\bibitem{Romero_2020}
E.~Romero, A.~Stathopoulos and K.~Orginos, \emph{Multigrid deflation for
  lattice qcd}, \href{https://doi.org/10.1016/j.jcp.2020.109356}{\emph{Journal
  of Computational Physics} {\bfseries 409} (2020) 109356}.

\bibitem{Brannick_2008}
J.~Brannick, R.C.~Brower, M.A.~Clark, J.C.~Osborn and C.~Rebbi, \emph{Adaptive
  multigrid algorithm for lattice qcd},
  \href{https://doi.org/10.1103/PhysRevLett.100.041601}{\emph{Phys. Rev. Lett.}
  {\bfseries 100} (2008) 041601}.

\bibitem{babich_2010}
R.~Babich, J.~Brannick, R.C.~Brower, M.A.~Clark, T.A.~Manteuffel,
  S.F.~McCormick et~al., \emph{Adaptive multigrid algorithm for the lattice
  wilson-dirac operator},
  \href{https://doi.org/10.1103/PhysRevLett.105.201602}{\emph{Phys. Rev. Lett.}
  {\bfseries 105} (2010) 201602}.

\bibitem{osborn_2010}
J.C.~Osborn, R.~Babich, J.~Brannick, R.C.~Brower, M.A.~Clark, S.D.~Cohen
  et~al., \emph{Multigrid solver for clover fermions},  LATTICE2010, 2010,
  \href{https://arxiv.org/abs/1011.2775}{https://arxiv.org/abs/1011.2775}
  [\href{https://arxiv.org/abs/1011.2775}{{\ttfamily 1011.2775}}].

\bibitem{Frommer_2014}
A.~Frommer, K.~Kahl, S.~Krieg, B.~Leder and M.~Rottmann, \emph{Adaptive
  aggregation-based domain decomposition multigrid for the lattice
  wilson--dirac operator}, \href{https://doi.org/10.1137/130919507}{\emph{SIAM
  Journal on Scientific Computing} {\bfseries 36} (2014) A1581}
  [\href{https://arxiv.org/abs/1303.1377}{{\ttfamily 1303.1377}}].

\bibitem{Richtmann_2022}
D.~Richtmann, N.~Meyer and T.~Wettig, \emph{Mrhs multigrid solver for
  wilson-clover fermions},  LATTICE2022, 2022,
  \href{https://arxiv.org/abs/2211.13719}{https://arxiv.org/abs/2211.13719}
  [\href{https://arxiv.org/abs/2211.13719}{{\ttfamily 2211.13719}}].

\bibitem{Cohen_2012}
S.D.~Cohen, R.C.~Brower, M.A.~Clark and J.C.~Osborn, \emph{Multigrid algorithms
  for domain-wall fermions},  LATTICE2012, 2012,
  \href{https://arxiv.org/abs/1205.2933}{https://arxiv.org/abs/1205.2933}
  [\href{https://arxiv.org/abs/1205.2933}{{\ttfamily 1205.2933}}].

\bibitem{boyle_2014}
P.A.~Boyle, \emph{Hierarchically deflated conjugate gradient},  2014.

\bibitem{Brower_2020}
R.C.~Brower, M.A.~Clark, E.~Weinberg and D.~Howarth, \emph{Multigrid for chiral
  lattice fermions: Domain wall},
  \href{https://doi.org/10.1103/PhysRevD.102.094517}{\emph{Phys. Rev. D}
  {\bfseries 102} (2020) 094517}.

\bibitem{Brower_2018}
R.C.~Brower, E.~Weinberg, M.A.~Clark and A.~Strelchenko, \emph{Multigrid
  algorithm for staggered lattice fermions},
  \href{https://doi.org/10.1103/PhysRevD.97.114513}{\emph{Phys. Rev. D}
  {\bfseries 97} (2018) 114513}.

\bibitem{Ayyar_2023}
V.~Ayyar, E.~Weinberg, R.C.~Brower, M.~Clark and M.~Wagner, \emph{Optimizing
  staggered multigrid for exascale performance},  in \emph{Proceedings of The
  39th International Symposium on Lattice Field Theory — PoS(LATTICE2022)},
  LATTICE2022, Sissa Medialab, Jan., 2023,
  \href{https://doi.org/10.22323/1.430.0335}{DOI}
  [\href{https://arxiv.org/abs/2212.12559}{{\ttfamily 2212.12559}}].

\bibitem{Clark_2010}
M.~Clark, R.~Babich, K.~Barros, R.~Brower and C.~Rebbi, \emph{Solving lattice
  qcd systems of equations using mixed precision solvers on gpus},
  \href{https://doi.org/10.1016/j.cpc.2010.05.002}{\emph{Computer Physics
  Communications} {\bfseries 181} (2010) 1517–1528}.

\bibitem{Babich_2011}
R.~Babich, M.A.~Clark, B.~Joó, G.~Shi, R.C.~Brower and S.~Gottlieb,
  \emph{Scaling lattice qcd beyond 100 gpus},  in \emph{Proceedings of 2011
  International Conference for High Performance Computing, Networking, Storage
  and Analysis}, SC ’11, p.~1–11, ACM, Nov., 2011,
  \href{https://doi.org/10.1145/2063384.2063478}{DOI}
  [\href{https://arxiv.org/abs/1109.2935}{{\ttfamily 1109.2935}}].

\bibitem{Stathopoulos_2009}
A.~Stathopoulos, A.M.~Abdel-Rehim and K.~Orginos, \emph{Deflation for inversion
  with multiple right-hand sides in qcd},
  \href{https://doi.org/10.1088/1742-6596/180/1/012073}{\emph{Journal of
  Physics: Conference Series} {\bfseries 180} (2009) 012073}.

\bibitem{Clark_2024}
K.~Clark, \emph{Rearchitecting quda for multi-rhs computations},  LATTICE2024,
  2024.

\bibitem{Weinberg_2024}
E.~Weinberg, \emph{Quda-accelerated batched solvers for lqcd workflows},
  LATTICE2024, 2024.

\bibitem{Clark_2018}
M.A.~Clark, C.~Jung and C.~Lehner, \emph{Multi-grid lanczos},
  \href{https://doi.org/10.1051/epjconf/201817514023}{\emph{EPJ Web of
  Conferences} {\bfseries 175} (2018) 14023}.

\end{thebibliography}\endgroup

\end{document}